\documentclass[notitlepage]{article}

\usepackage{natbib}
\usepackage{titlesec}
\usepackage[titletoc,toc,title]{appendix}
\titleformat{\subsection}{\normalfont\itshape}{\thesubsection.}{0.5em}{}
\titleformat{\subsubsection}{\normalfont\itshape}{\thesubsubsection.}{0.6em}{}
\usepackage{adjustbox}
\usepackage{booktabs,siunitx,rotating}
\usepackage{lineno}
\usepackage{pgf,tikz}
\usepackage{appendix}
\usepackage{mathrsfs}
\usepackage{amsmath}
\usepackage{multirow}
\usepackage{booktabs} 
\newcommand{\ra}[1]{\renewcommand{\arraystretch}{#1}}
\usepackage{mathtools}
\usepackage{array}
\usetikzlibrary{arrows}
\usepackage{pdflscape}
\usepackage{textcmds}
\usepackage{amssymb}
\usepackage{hyperref}
\usepackage{textcmds}
\usepackage{caption}
\usepackage[nottoc,notlot,notlof]{tocbibind}

\usepackage{etoolbox}
\usepackage{lipsum} 
\usepackage[utf8]{inputenc} % allow utf-8 input
\usepackage[T1]{fontenc}    % use 8-bit T1 fonts
\usepackage{hyperref}       % hyperlinks
\usepackage{url}            % simple URL typesetting
\usepackage{booktabs}       % professional-quality tables
\usepackage{amsfonts}       % blackboard math symbols
\usepackage{nicefrac}       % compact symbols for 1/2, etc.
\usepackage{microtype}      % microtypography
\usepackage{sidecap}
\usepackage{lscape}
\usepackage{wrapfig}
\usepackage{multirow}
\usepackage{enumitem}
\usepackage{graphicx}
\usepackage{dcolumn}
\usepackage{subcaption}
\usepackage{indentfirst}
\usepackage{amsmath, amssymb, mathtools, bm, amsthm}
\usepackage{amsfonts}
\usepackage{gensymb}
\usepackage{placeins}
\usepackage{booktabs} % for professional tables
\usepackage{todonotes}
\usepackage{hyperref}
\usepackage[ruled,vlined]{algorithm2e}
\usepackage{algorithmic}
\usepackage{bm}
\usepackage{cleveref}

\usepackage{pifont}% http://ctan.org/pkg/pifont
\title{Machine Learning, Deep Learning, and Hedonic Methods for Real Estate Price Prediction}

\author{\stepcounter{footnote} Mahdieh Yazdani\thanks{\noindent Department of Economics, Colorado University at Boulder. Email contact: mahdieh.yazdani@colorado.edu}}

\begin{document}

\begingroup
\maketitle
\thispagestyle{empty}

\begin{abstract}

In recent years several complaints about racial discrimination in appraising home values have been accumulating. For several decades, to estimate the sale price of the residential properties, appraisers have been walking through the properties, observing the property, collecting data, and making use of the hedonic pricing models. However, this method bears some costs and by nature is subjective and biased. To minimize human involvement and the biases in the real estate appraisals and boost the accuracy of the real estate market price prediction models, in this research we design data-efficient learning machines capable of learning and extracting the relation or patterns between the inputs (features for the house) and output (value of the houses). We compare the performance of some machine learning and deep learning algorithms, specifically artificial neural networks, random forest, and $k$ nearest neighbor approaches to that of hedonic method on house price prediction in the city of Boulder, Colorado. Even though this study has been done over the houses in the city of Boulder it can be generalized to the housing market in any cities. The results indicate non-linear association between the dwelling features and dwelling prices. In light of these findings, this study demonstrates that random forest and artificial neural networks algorithms can be better alternatives over the hedonic regression analysis for prediction of the house prices in the city of Boulder, Colorado. \\

% develops dwelling price prediction methods based on the machine learning and deep learning algorithms for the houses in the city of Boulder in the state of Colorado, USA. Even though this study has been done over the houses in the city of Boulder it can be generalized to the housing market in any cities. 

\noindent \textit{keywords}: Housing price prediction model, Artificial neural network, Random Forest, K nearest neighbor, regression analysis.\\

\end{abstract}
\thispagestyle{empty}
\newpage
\tableofcontents
\thispagestyle{empty}
\endgroup  

\clearpage
\setcounter{page}{1}
\section{Introduction}

In most families dwelling is one of the most important components of a household's wealth (see e.g., \cite{arvanitidis2014economics}). Consequently, house prices are of great interest to actual and potential homeowners. The property prices are not only important to stakeholders, but also to insurance companies, property developers, appraisers, tax assessors, brokers, banks, mortgage lenders, and policy makers (see e.g., \cite{frew2003estimating}). The price of residential properties is affected directly by their attributes such as structural factors, socio-economic status of the neighborhood, environmental amenities, and location. For several decades, the estimation of real estate assets have relied on hedonic pricing (HP) models. In this method, to estimate the sale price of the residential properties, appraisers usually walk through the properties, observe the house, collect data, and make use of the HP models. However, this method bears some costs and by nature is subjective and biased. Biased real estate appraisal is a common and longstanding issue in the United States. For example, in the recent years several complaints about racial discrimination in estimating home values have been accumulating (see e.g., \href{https://www.bloomberg.com/news/articles/2021-03-03/appraisers-acknowledge-bias-in-home-valuations}{Bloomberg CityLab}). Consequently, to obtain a fair home assessment and avoid subjectivity and biases in the dwelling appraisals, the need for 
development of accurate housing valuation and property price prediction models has been growing (see e.g., \cite{fan2018house}).\\

As alternative tools, machine learning and deep learning algorithms such as artificial neural networks (ANN) , $k$ nearest neighbor (kNN), bounded fuzzy possibilistic method (BFPM), random forest (RF), and support vector regression (SVR) have been offered for the house price valuation and real estate property price prediction model. Some studies such as \cite{bigus1996data}, \cite{lenk1997high}, \cite{kauko2002capturing}, \cite{curry2002neural},  \cite{pagourtzi2003real}, \cite{limsombunchai2004house}, \cite{peterson2009neural}, \cite{islam2009housing}, \cite{selim2011determinants}, \cite{morano2013bare}, \cite{ali2015housing}, \cite{park2015using}, \cite{yazdani2018comparative}, \cite{vceh2018estimating}, \cite{poursaeed2018vision},  \cite{hong2020house}, and \cite{pai2020using} utilized deep learning and machine learning techniques on real estate markets. Most of these studies found that deep learning and machine learning algorithms perform better than a hedonic regression method for real estate market appraisal. Namely, \cite{selim2011determinants} compared the prediction performances of HP regression and ANN model for prediction of dwelling prices in Turkey. It demonstrated that an ANN performs better. \cite{yao2018mapping} integrated a convolutional neural network with RF inside Shenzhen's housing market. \cite{park2015using} has compared the performance among several machine learning algorithms such as Repeated Incremental Pruning to Produce Error Reduction, Na{\"i}ve Bayesian, and AdaBoost to identify better forecasting models. These studies have shown promising application of machine learning and deep learning algorithms in housing markets. \\

In order to minimize human involvement and the biases in the real estate appraisals and boost the accuracy of the real estate market price prediction models, in this paper we design data-efficient learning machines capable of learning and extracting the relation or patterns between the inputs (features for the house) and output (value of the houses). In this study we compare the performance of some machine learning and deep learning algorithms, specifically ANN, RF, and kNN approaches to that of HP method on house price prediction to develop property appraisal models in the city of Boulder, Colorado. In the city of Boulder, a popular mountain city, buying a house is considered as one of the most profitable investments and most people in Boulder know the benefit of owning a house. Even though this study has been done over the houses in the city of Boulder it can be generalized to the housing market in any cities.\\

The objective of this study is to compare the predictive power of HP regression methods with machine learning and deep learning algorithms on house price prediction. Development of an accurate dwelling price prediction method can significantly improve the efficiency of the housing market. Having access to efficient and accurate house price prediction models can be important for actual and potential homeowners, investors, property developers, appraisers, tax assessors, brokers, banks, mortgage lenders, and policy makers and affect their decisions and policies (see e.g., \cite{frew2003estimating} and \cite{yazdani2020individuals}).\\

The contributions of this paper are as follows:\\

\begin{itemize} [leftmargin=*]
 \item  Collecting real estate datasets from different resources consisting of Multiple Listing Service (MLS) database, Public School Ratings, Colorado Crime Rates and Statistics Information, CrimeReports, WalkScore, and US Census Bureau. 
\end{itemize} 

\begin{itemize} [leftmargin=*]

 \item Screening the collected data set, data cleansing, and applying imputation and winsorization methods.  
\end{itemize} 

\begin{itemize} [leftmargin=*]

 \item Detecting multicollinearity problems, standardizing the numerical variables, and applying the one-hot encoding method over the categorical variables.
\end{itemize}

\begin{itemize} [leftmargin=*]
\item Applying the ANN, RF, and kNN approaches to develop property appraisal models and comparing these methods with HP model.
\end{itemize} 

The rest of the paper is organized as follows. Section 2 provides an overview of the hedonic price, deep learning, and machine learning approaches. Following this, section 3 explains the empirical models and presents the data set, variables and methodology used in this paper. The estimation results obtained by these techniques described in section 4. Finally, conclusions and implications are provided in section 5.\\

\section{Modeling the Housing Market}

We collected a labelled data set composed of input and target variables $(X,Y)$, where in this study the residential property features are the input variables $X$ and the vector $Y$ is the output variable consisting of logarithm of house values. Our goal is therefore to learn a function

\begin{equation}
F: X \rightarrow Y 
\end{equation}\\

\noindent such that given new residential property features $X$, $F(X)$ can predict the house value $Y$.\\

In this study we first briefly introduce HP, ANN, RF, and kNN techniques, and next apply these methods to our data set and compare their performance.\\

\subsection{Hedonic Pricing Model}

Hedonic price analysis originates from \cite{waugh1928quality}, \cite{at1939hedonic}, \cite{stone1954measurement}, \cite{adelman1961index}, and \cite{lancaster1966new}. The literature on HP methodology is broad. For example, \cite{rosen1974hedonic} used the hedonic regression models in the field of real estate and urban economics. Thereafter, residential hedonic analysis has become widely used as an assessment tool to model real estate markets (see e.g., \cite{mcmillan1980extension}, \cite{blomquist1981hedonic}, \cite{milon1984hedonic}, \cite{haase2013tools}, \cite{jiang2014new}, \cite{del2017hedonic}, \cite{krol2020application}, and \cite{yazdani2021house}). In the housing market the HP regression model is essentially a function of a bundle of property characteristics such as structural attributes, socio-economic status of the neighborhood, environmental amenities, and location. This model is constructed by regressing the price of the dwellings on a vector of housing features. The HP models have been used to explain variations in dwelling prices and to determine the impact of specific attributes on house values (see e.g., \cite{geng2015study}). By differentiating the HP model with respect to each feature we can easily find the marginal contribution or implicit valuations of the house features  (see \cite{mcmillan1980extension}). The simplicity in estimating and interpreting the regression coefficients is the major advantage of the HP models.  \\

Even though the HP regression approach has been extensively used to analyze the market price of houses, it is limited in its ability to uncover the non-linear relationships between housing price and its characteristics. Factors such as model specification procedures, heteroscedasticity, and outlier data points are other issues that can impede the performance of hedonic regression models. In general, HP method needs feature engineering which involves human and consequently, subjectivity and biases in the real estate appraisals. There are, however, some alternative techniques, which are better suited to deal with these problems. As alternative tools, machine learning and deep learning algorithms are more flexible than the hedonic method. Unlike the HP regression model, there are neither strict assumptions nor predefined functional form to be employed in the deep learning and machine learning algorithms and we do not need to specify an explicit function between input and target variables. Instead, these algorithms learn from the observed data set and explore the linear and non-linear relationship between attributes of the properties and the property values. Some studies such as (\cite{lenk1997high}, \cite{kauko2002capturing}, \cite{curry2002neural},  \cite{pagourtzi2003real}, \cite{limsombunchai2004house}, \cite{yu2006incorporating}, \cite{islam2009housing}, \cite{selim2011determinants}, \cite{kontrimas2011mass}, \cite{morano2013bare}, \cite{ali2015housing}, \cite{park2015using}, \cite{vceh2018estimating}, and \cite{hong2020house}) utilized deep learning and machine learning techniques on real estate markets to develop housing price prediction models. Most of these studies found that deep learning and machine learning algorithms perform better than a hedonic regression method for estimating dwelling values.\\ 

\subsection{Artificial Neural Network Model}

The first artificial neural network was designed by psychologist Frank Rosenblatt (see \cite{rosenblatt1958perceptron}). Artificial neural network is a deep learning technique that was inspired by the human brain (see e.g., \cite{bengio2021deep}). ANN is designed to simulate the learning process in the human brain (see e.g., \cite{collins1998design}, \cite{agatonovic2000basic}, \cite{vazirizade2017seismic}, and \cite{fleuret2020deep}). In particular, the objective of designing a neural network is to learn the association between inputs and outputs in the given data set, so that for any new inputs, the corresponding output could be determined. \\

\begin{figure}[h!]
\centering
\hspace*{.08\textwidth} 
\includegraphics[width=0.7\textwidth]{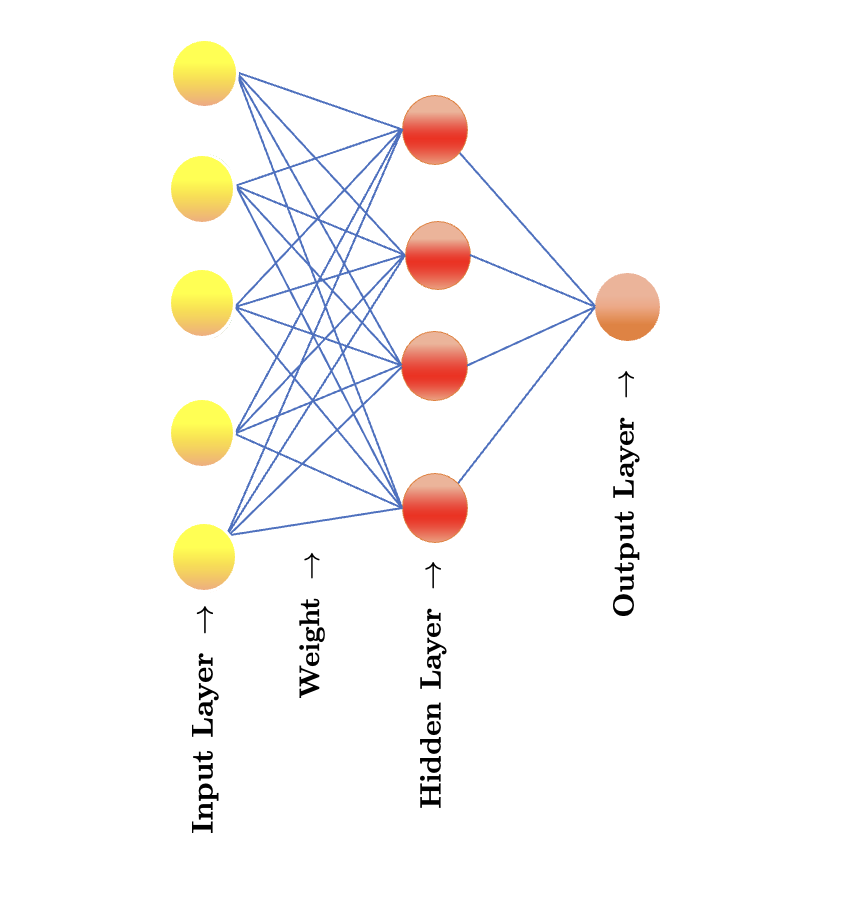}
\caption{An Artificial Neural Network Model. }\label{Figure1}
\end{figure}

In ANN there is no need to assume explicit functions between input and output of the studies. ANN learns directly from observed data and models non-linear and compound relationships. An ANN requires a number of processing units, called neurons (nodes), to be connected together into a neural network (see e.g., \cite{heiat2002comparison} and \cite{yegnanarayana2009artificial}). The ANN method consists of multiple layers: input, hidden, and output layers. Each layer contains one or more neurons. Every neuron in one layer is connected to each single node at the neighbor layer(s), and each connection is associated with a weight, which determines how strongly each of the neurons affects the others (see e.g., \cite{haykin2004comprehensive} and \cite{kassambara2018machine}). The number of neurons and layers in different artificial neural networks can vary. The number of neurons in the hidden layers determines the width of the model and the number of layers determines the depth of the model. Figure \ref{Figure1} exhibits a simplified feedforward neural network model. This network consists of three layers of neurons in which each neuron is connected to all the neurons in the next layer. The input enters the network through the input layer, it is multiplied by a weight matrix and the result passes through an activation function (transfer function) to the intermediate hidden layer, which is located between the input and output layer. An activation function defines how the weighted sum of the input is transformed into an output from a neuron or neurons in a layer of the network. Usually, the purpose of an activation function is to introduce non-linearity to help identify the underlying patterns which are complex (see e.g., \cite{hopfield1982neural}). Similarly, 
the hidden layer through an activation function performs linear or non-linear transformations over the output of the previous layer (the input layer in this figure) and directs them to the output layer. Finally, the output layer performs the calculations and then the outputs or solutions are computed. \\

In an ANN, during the learning process, the values of the weight vector are adjusted to minimize the loss function. The most often loss function used in a supervised ANN is defined as\\

$L= {\frac {\sum _{i=1}^n{(y_i -\hat{y_i})}^2}{n}}$\\

\noindent where L is the mean squared errors, where error is the difference between actual output ${y_i}$ and estimated output $\hat{y_i}$ for a given input and $n$ is the number of samples.\\

To minimize the loss function all of the connection weights in the network are gradually adjusted, using the backpropagation method to do gradient descent optimization. In backpropagation we start from the output layer through the hidden layer(s), and to the input layer, by adjusting weights for each layer throughout the network. Computing the sensitivity of a loss function with respect to each weight in the network, will eventually lead to minimization of error between the predicted and actual outputs. We update each weight until we get estimated outputs which are close to the actual outputs for a particular input. This process of adjusting the weights is referred to as learning. In the learning process the weights of the connecting links in a network are adjusted so as to enable the network to learn the pattern between the given input and output data set. As the learning process proceeds, the connections which lead to the correct solutions are strengthened, and the incorrect connections are weakened. \\

In real life almost all of our data is non-linear. ANNs are usually efficient in predicting complex systems that demonstrate non-linear behavior. ANN has applications in many different fields, namely medical diagnosis, image processing, voice recognition, credit scoring, evaluating loan applications, forecasting, fraud detection, and making decisions. By changing the activation functions of the output layer, ANN can be applied to both classification and regression problems.\\

In this study, input and output variables in the ANN are the same as in the HP regression model. The inputs are features for the house, and the predicted outputs are the estimated price of the houses. In this study we employ the feed forward network to estimate the sale price of the residential properties. We use the Rectified Linear Unit (ReLU) as the activation function (see e.g., \cite{lecun1998gradient}). The ReLU activation function is easy to optimize because it is similar to linear function. A ReLU activation function will output the input directly if it is positive, otherwise, it will output zero.\\ 

%The importance of the activation function is that it normally introduces non-linearities into the data. 

\subsection{Random Forest Model}

Random forest, a supervised learning technique, is one of the most robust machine learning algorithms that can be used for both classification and regression problems (see \cite{breiman2001random}). RF is a special type of bootstrap aggregation called bagging that combines many decision trees (see e.g., \cite{breiman1996bagging} and \cite{ho1998random}). The idea is to train several different trees separately, and then integrate the predictions from multiple decision trees to make more reliable predictions than that of any single decision tree. In RF each decision tree is grown independently on a separate bootstrap sample derived from the training data set. Figure \ref{Figure 2} illustrates an example of RF. \\

In general, to make the RF predictions more reliable, more trees are required. However, increasing the number of trees slows down the prediction process. To limit overfitting and avoid high correlation amongst the decision trees the RF technique uses bagging (see e.g., \cite{breiman1996bagging}). In addition, each decision tree is grown on only a random selection of attributes (predictors). To construct each single decision tree, increasing the number of variables, when deciding to split a node, increases both the internal correlation and the predicted power. Reducing it decreases both (see e.g., \cite{strobl2008conditional}). Somewhere in between is an optimal range for the number of variables. The optimal number of features (also known as mtry in R) is normally around the square root of the number of features in the classification method. In the regression approach, the optimal number of variables is usually around the $\frac{1}{3}$ of the number of characteristics. Reducing the maximum depth will also help reduce the complexity of the model and avoid overfitting.\\

\begin{landscape}

\begin{figure}
\centering
\begin{tikzpicture}[node distance=2cm]
%\draw [draw=none, fill=green] (-30:1.5)

\node (A) at (-8.5, 8) [rectangle,draw,fill=cyan,rounded corners=5pt] {Statement};
\node (B1) at (-9.85, 6.5) [rectangle,draw,fill=green!40!yellow,rounded corners=5pt] {Statement};
\node (B2) at (-7.3, 6.5) [rectangle,draw,fill=green!40!yellow,rounded corners=5pt]{Statement};

\node (C1) at (-11,5) [rectangle,draw,fill=green!40!yellow,rounded corners=5pt] {Statement};
\node (C2) at (-9,5)[rectangle,draw,fill=orange!40!yellow,rounded corners=5pt] {Leaf};
\node (C3) at (-7.6,5) [rectangle,draw,fill=orange!50!yellow,rounded corners=5pt]{Leaf};
\node (C4) at (-5.8,5) [rectangle,draw,fill=green!40!yellow,rounded corners=5pt]{Statement};

%%%%

\draw[thick,->] (A) -- node[left] {\textit{True}} (B1);
\draw[thick,->] (A) -- node[right] {\textit{False}} (B2);
\draw[thick,->] (B1) -- node[left] {\textit{True}} (C1);
\draw[thick,->] (B1) -- node[right] {\textit{False}} (C2);
\draw[thick,->] (B2) -- node[left] {\textit{True}} (C3);
\draw[thick,->] (B2) -- node[right] {\textit{False}} (C4);

%Text\myline Text\myline[dashed]Text\myline[dotted]

\draw[thick,densely dashed](-4.2,6.5) -- (-2.5,6.5);

%%%%%%%% The figure on the right! Shifted by 8 points! %%%%%%%%%%%%
\node (Ap) at (1.5, 8) [rectangle,draw,fill=cyan,rounded corners=5pt] {Statement};
\node (Bp1) at (.3, 6.5) [rectangle,draw,fill=green!40!yellow,rounded corners=5pt]{Statement};
\node (Bp2) at (2.85, 6.5) [rectangle,draw,fill=green!40!yellow,rounded corners=5pt]{Statement};

\node (Cp1) at (-1.6,5) [rectangle,draw,fill=orange!50!yellow,rounded corners=5pt] {Leaf};
\node (Cp2) at (.3,5) [rectangle,draw,fill=green!40!yellow,rounded corners=5pt]{Statement};
\node (Cp3) at (2.8,5) [rectangle,draw,fill=orange!50!yellow,rounded corners=5pt] {Leaf};
\node (Cp4) at (4.8,5) 
[rectangle,draw,fill=green!40!yellow,rounded corners=5pt]{Statement};
%%%%

\draw[thick,->] (Ap) -- node[left] {True} (Bp1);
\draw[thick,->] (Ap) -- node[right] {False} (Bp2);
\draw[thick,->] (Bp1) -- node[left] {True} (Cp1);
\draw[thick,->] (Bp1) -- node[right] {False} (Cp2);
\draw[thick,->] (Bp2) -- node[left] {True} (Cp3);
\draw[thick,->] (Bp2) -- node[right] {False} (Cp4);

%%%%%%%%%%%%
\node (App1) at (-12.5, 3.5) [rectangle,draw,fill=orange!50!yellow,rounded corners=5pt] {Leaf};

\node (App2) at (-11, 3.5) [rectangle,draw,fill=orange!50!yellow,rounded corners=5pt] {Leaf};

%%%%

\draw[thick,->] (C1) -- node[left] {\textit{True}} (App1);
\draw[thick,->] (C1) -- node[right] {\textit{False}} (App2);

%%%%%%%%%%%%
\node (App3) at (-7, 3.5) [rectangle,draw,fill=orange!50!yellow,rounded corners=5pt] {Leaf};

\node (App4) at (-5.5, 3.5) [rectangle,draw,fill=orange!50!yellow,rounded corners=5pt] {Leaf};

%%%%

\draw[thick,->] (C4) -- node[left] {\textit{True}} (App3);
\draw[thick,->] (C4) -- node[right] {\textit{False}} (App4);

%%%%%%%%%%%%
\node (Bpp1) at (-1, 3.5) [rectangle,draw,fill=orange!50!yellow,rounded corners=5pt] {Leaf};

\node (Bpp2) at (1, 3.5) [rectangle,draw,fill=orange!50!yellow,rounded corners=5pt] {Leaf};

%%%%

\draw[thick,->] (Cp2) -- node[left] {\textit{True}} (Bpp1);
\draw[thick,->] (Cp2) -- node[right] {\textit{False}} (Bpp2);

%%%%%%%%%%%%
\node (Bpp3) at (3.5, 3.5) [rectangle,draw,fill=orange!50!yellow,rounded corners=5pt] {Leaf};

\node (Bpp4) at (5.5, 3.5) [rectangle,draw,fill=orange!50!yellow,rounded corners=5pt] {Leaf};

%%%%

\draw[thick,->] (Cp4) -- node[left] {\textit{True}} (Bpp3);
\draw[thick,->] (Cp4) -- node[right] {\textit{False}} (Bpp4);

%%%%

\draw[thick,densely dashed](-8.5,3) -- (-8.5,1.5);

\draw[thick,densely dashed](1.75,3) -- (1.75,1.5);
%%%%%%%%%%%%
\node (A) at (-8.5, 0) [rectangle,draw,fill=cyan,rounded corners=5pt] {Statement};
\node (B1) at (-9.85, -1.5) [rectangle,draw,fill=green!40!yellow,rounded corners=5pt] {Statement};
\node (B2) at (-7.3, -1.5) [rectangle,draw,fill=green!40!yellow,rounded corners=5pt] {Statement};

\node (C1) at (-11,-3) [rectangle,draw,fill=green!40!yellow,rounded corners=5pt] {Statement};
\node (C2) at (-9,-3)[rectangle,draw,fill=orange!40!yellow,rounded corners=5pt] {Leaf};
\node (C3) at (-7.6,-3) [rectangle,draw,fill=orange!50!yellow,rounded corners=5pt]{Leaf};
\node (C4) at (-5.8,-3) [rectangle,draw,fill=orange!40!yellow,rounded corners=5pt] {Leaf};

%%%%

\draw[thick,->] (A) -- node[left] {\textit{True}} (B1);
\draw[thick,->] (A) -- node[right] {\textit{False}} (B2);
\draw[thick,->] (B1) -- node[left] {\textit{True}} (C1);
\draw[thick,->] (B1) -- node[right] {\textit{False}} (C2);
\draw[thick,->] (B2) -- node[left] {\textit{True}} (C3);
\draw[thick,->] (B2) -- node[right] {\textit{False}} (C4);

\draw[thick,densely dashed](-4.2,-1.5) -- (-2.5,-1.5);

%%%%%%%% The figure on the right! Shifted by 8 points! %%%%%%%%%%%%
\node (Ap) at (1.5, 0) [rectangle,draw,fill=cyan,rounded corners=5pt] {Statement};
\node (Bp1) at (.3, -1.5) [rectangle,draw,fill=green!40!yellow,rounded corners=5pt]{Statement};
\node (Bp2) at (2.85, -1.5) [rectangle,draw,fill=green!40!yellow,rounded corners=5pt]{Statement};

\node (Cp1) at (-1.6,-3)[rectangle,draw,fill=green!40!yellow,rounded corners=5pt]{Statement};
\node (Cp2) at (.3,-3) [rectangle,draw,fill=orange!50!yellow,rounded corners=5pt] {Leaf};
\node (Cp3) at (2.8,-3) [rectangle,draw,fill=green!40!yellow,rounded corners=5pt]{Statement};
\node (Cp4) at (4.8,-3) [rectangle,draw,fill=orange!50!yellow,rounded corners=5pt] {Leaf};

%%%%

\draw[thick,->] (Ap) -- node[left] {True} (Bp1);
\draw[thick,->] (Ap) -- node[right] {False} (Bp2);
\draw[thick,->] (Bp1) -- node[left] {True} (Cp1);
\draw[thick,->] (Bp1) -- node[right] {False} (Cp2);
\draw[thick,->] (Bp2) -- node[left] {True} (Cp3);
\draw[thick,->] (Bp2) -- node[right] {False} (Cp4);

%%%%%%%%%%%%
\node (App1) at (-12.5, -4.5) [rectangle,draw,fill=orange!50!yellow,rounded corners=5pt] {Leaf};

\node (App2) at (-11, -4.5) [rectangle,draw,fill=orange!50!yellow,rounded corners=5pt] {Leaf};

%%%%

\draw[thick,->] (C1) -- node[left] {\textit{True}} (App1);
\draw[thick,->] (C1) -- node[right] {\textit{False}} (App2);

%%%%%%%%%%%%
\node (Appp1) at (-1.6, -4.5) [rectangle,draw,fill=orange!50!yellow,rounded corners=5pt] {Leaf};

\node (Appp2) at (0, -4.5) [rectangle,draw,fill=orange!50!yellow,rounded corners=5pt] {Leaf};

%%%%

\draw[thick,->] (Cp1) -- node[left] {\textit{True}} (Appp1);
\draw[thick,->] (Cp1) -- node[right] {\textit{False}} (Appp2);

%%%%%%%%%%%%
\node (Appp3) at (2.8, -4.5) [rectangle,draw,fill=orange!50!yellow,rounded corners=5pt] {Leaf};

\node (Appp4) at (4.5, -4.5) [rectangle,draw,fill=orange!50!yellow,rounded corners=5pt] {Leaf};

%%%%

\draw[thick,->] (Cp3) -- node[left] {\textit{True}} (Appp3);
\draw[thick,->] (Cp3) -- node[right] {\textit{False}} (Appp4);
\end{tikzpicture}
\caption{A Random Forest Model.}\label{Figure 2}
\end{figure}
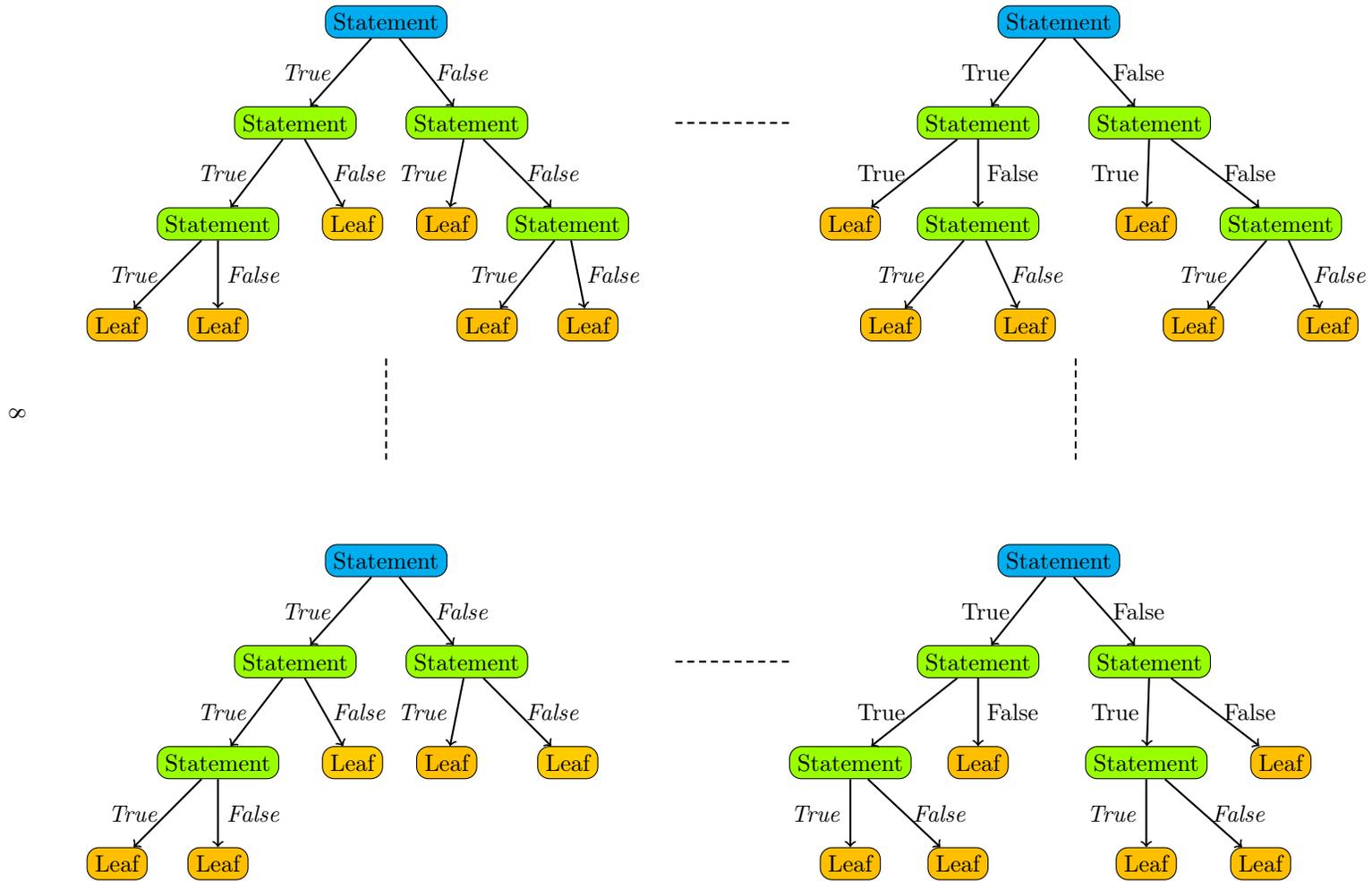
\end{landscape}

RF is an ensemble technique (see e.g., \cite{breiman2001random}) that integrates the predictions from multiple decision trees to make more reliable predictions than that of any single decision tree. After a large number of trees are generated, depending on the task, whether it is classification or regression problem RF aggregates the results from the decision trees by voting or averaging respectively. In the classification problem the RF method gives a prediction based on a majority vote. The random forest regression model produces the final prediction by averaging the predictions made by each of its decision trees (see e.g., \cite{antipov2012mass}). Merging decision trees' predictions, leads to more robust overall predictions and the predictions will be closer to the true value on average.\\

RF has many advantages. It is relatively easy to build and does not require too much computational time and the hyperparameters tuning is not too expensive. Since the RF algorithm is based on the bootstrapping, it performs well on even small data sets. Moreover, RF has few, if any, statistical assumptions. It does not assume that data is normally distributed, the relationship is linear, or there are specified interactions. Interpretability of the trained model is another major advantage of the RF method. Additionally, it is very easy to measure the relative importance of each feature on the output variable.\\

The RF algorithm is used in many different fields, namely stock market, banking, health care, and housing market. A number of researchers applied the RF technique in the real estate market (see e.g., \cite{antipov2012mass}, \cite{yoo2012variable}, \cite{hong2020house}, and \cite{levantesi2020importance}). \\

\subsection{K Nearest Neighbors Model}

Another machine learning method for housing market valuation is the k nearest neighbors approach. kNN is a supervised learning technique that has been used in both regression and classification problems. The kNN method uses attributes similarity to predict the values of any new data points. It checks how alike is a data point (a vector of features) in the test data set from other available data points in the training data set. The kNN method is similar to what the appraisers do in the real estate market. To determine a house price, realtors or appraisers search for dwellings with akin features, namely square footage, number of rooms, property type, location, neighborhood, and socio-economic attributes. They would value the new house around the average price that these properties sold for. Likewise, in the kNN approach the estimated property value is the average of the values of its $k$ nearest neighbors (see e.g., \cite{kassambara2018machine}).\\

KNN is a distance based model. In both regression and classification problems kNN approach generally measures the similarity between the available transactions and the new house by adopting a distance function. There are various methods such as Euclidean, Manhattan, Minkowski, Chebychev, Mahalanobis, and Hamming distance that we could use, to measure the distance between the target point and each training point and calculate the alikeness of data points based on all factors. Once the distance of a new observation from the current observations has been measured kNN finds the $k$ data points that are alike or nearest to the new record. In the kNN regression approach, the value of the new numerical data point is predicted by calculating the average of the observations in the $k$ closest neighborhood. The average of the $k$ nearest neighbors is the prediction for the numerical target. In contrast to the kNN regression method, in kNN classification problem, to classify a new data point in the test data set, kNN looks for its $k$ nearest points in the training samples and predicts the class of the target point in the test sample based on the mode of its $k$ nearest neighbors or the majority of classes among the $k$ closest imminent. It then assigns the corresponding label to the observation (see e.g., \cite{larose2005k} and \cite{domeniconi2002locally}).\\%In the kNN regression approach, the relationship between independent variables and the numerical target are predicted by calculating the average of the observations in the $K$ immanent neighborhood.

% and actually there is not a training stage or learning process. 

% kNN is a lazy learning or an instance-based method (see e.g., \cite{aha2013lazy}). This means kNN does not explicitly learn a model. kNN memorizes the training data set and stores the training data set to use for the prediction. 

KNN method is an efficient and accurate technique that can be used in a variety of applications such as finance, economic forecasting or prediction, climate forecasting, text mining, and genetics (see e.g., \cite{larose2014discovering}). One major advantage of the kNN approach is that it is a non-parametric method, which means there are no assumptions about how the data set has been distributed. Consequently, if we do not have enough information about the distribution of our data, the kNN method might be the right model to pick (see e.g., \cite{devroye1994strong} and \cite{devroye19828}). \\

kNN is a lazy learning or an instance-based method (see e.g., \cite{aha2013lazy}). This means kNN does not explicitly learn a model. kNN memorizes the training data set and stores the training data set for the prediction. When a new data point is given the kNN algorithm uses the training data set for the prediction. For every point in the test data set, the model has to compute the distance between each data point in the test data set and all data points in the training data set and then find the nearest neighbors. Next, it returns the average of the corresponding output values in the training set. Therefore, it needs high memory to store all this information. As such, the computation and memory costs are quite high. When there are many independent variables the kNN method becomes impractical. In addition, kNN is not robust against irrelevant factors since in distance metrics all factors contribute to the similarity. However, by careful feature selection or weighting the feature, this can be avoided.\\

\begin{figure}[h!]
\centering
\hspace*{.01\textwidth} 
\includegraphics[width=0.7\textwidth]{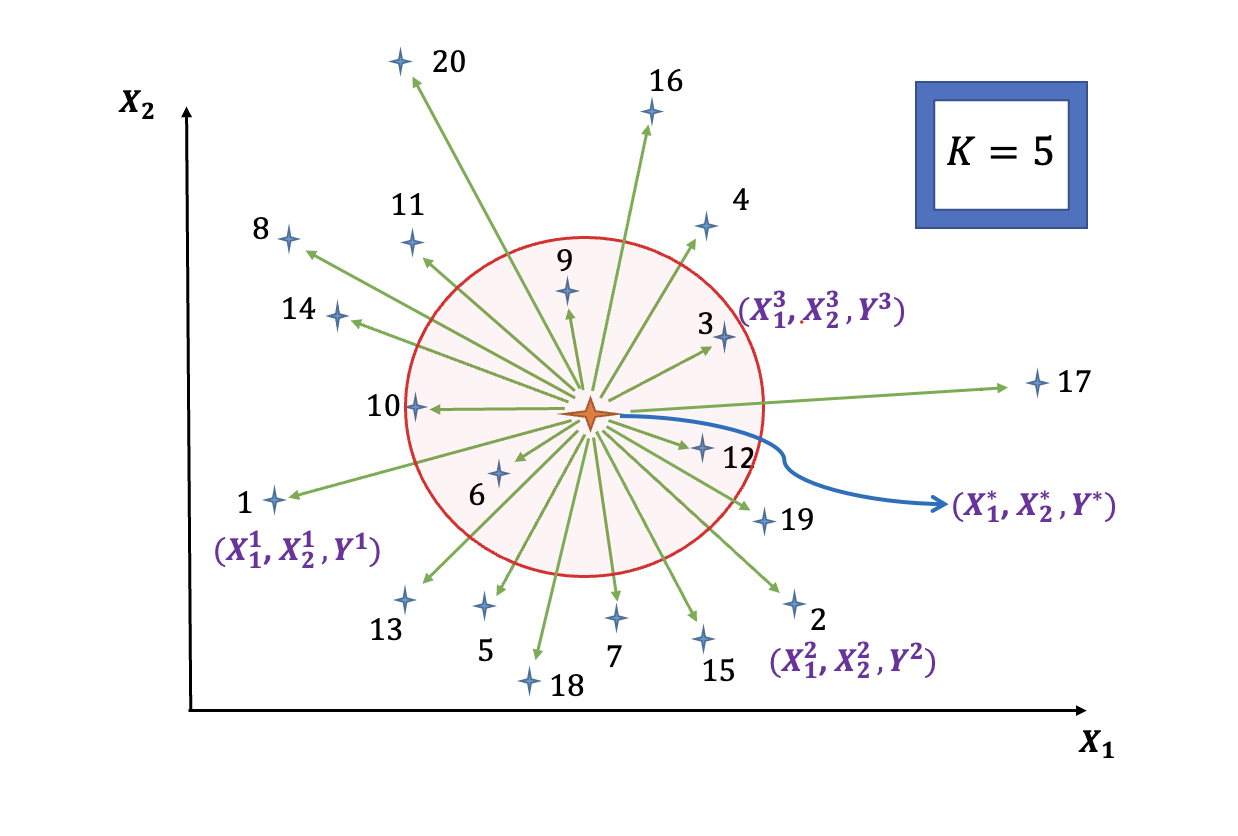}
\caption{A K Nearest Neighbors Model. }\label{Figure3}
\end{figure}
 
In this study the target variable is the house value. Since the house value is a continuous variable we will focus on the kNN regression approach. Figure \ref{Figure3} shows a kNN algorithm for $k=5$. The predicted outcome ($\widehat{Y^*}$) for the new data point (the star with the red color) would be the average of the outcomes of its $5$ closest neighbors.\\

This paper examines the relative influence of property characteristics, socio-economic factors, and the impact of accessibility on the price of residential properties in the city of Boulder, Colorado. In this study we use HP analysis, ANN, RF, and kNN approaches, to develop housing price prediction models in the city of Boulder, Colorado. Before proceeding with these models, we are interested in getting some insights about the available data set. \\

\section{Data and Functional Forms}

\begin{figure}[h!]
\centering
\hspace*{.00001\textwidth} 
\includegraphics[width=0.95\textwidth ,height=0.45\textheight]{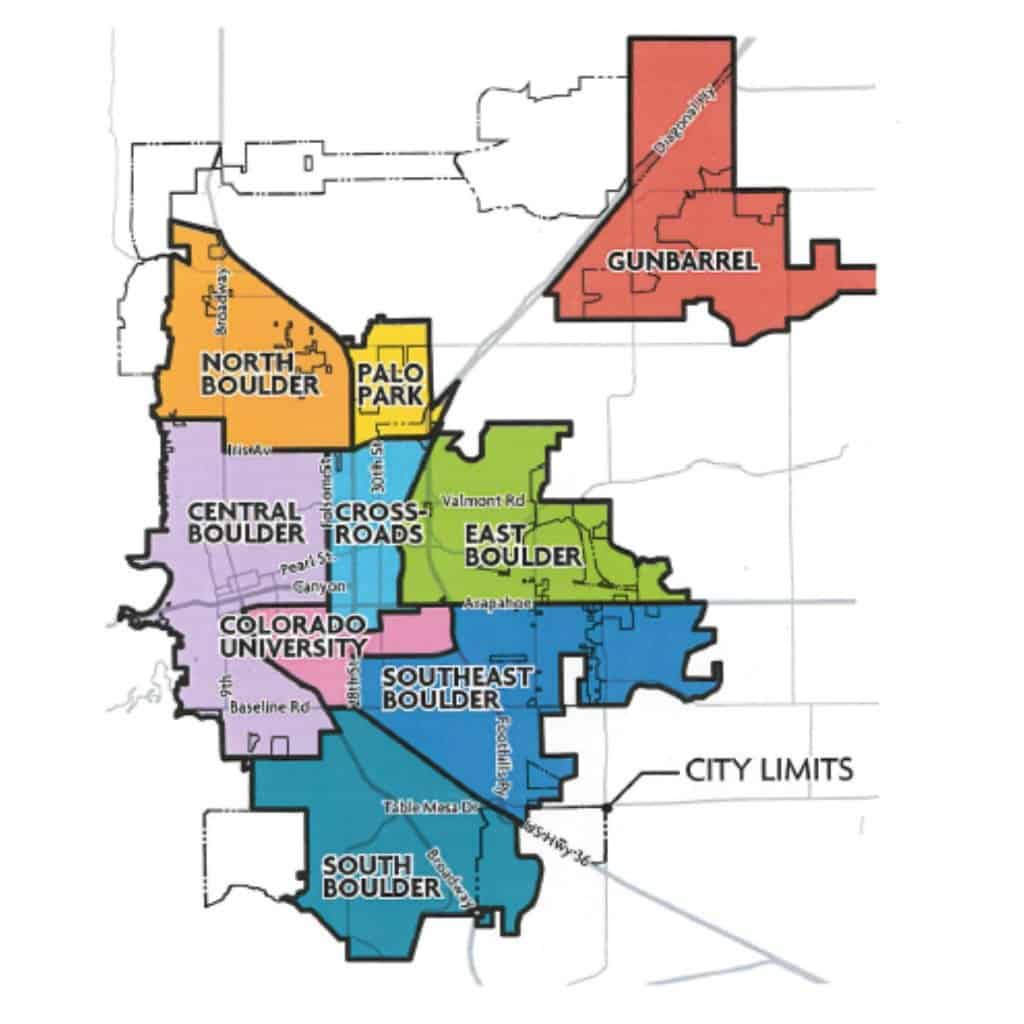}
\caption{City of Boulder.}\label{Figure4}
\end{figure}

In general, machine learning and deep learning techniques are used over large data sets to recognize hidden patterns that are difficult to detect through regression models. However, it is also applicable in small data sets (see e.g., \cite{steffensen2021risks} and \cite{levantesi2020importance}). The data set we collected in this study consists of $1061$ residential properties sold in the city of Boulder, Colorado. Figure \ref{Figure4} plots the city of Boulder on the map. To isolate the influence of time on property prices, the data used in this study is restricted to houses sold in a single year between January 1, 2019 and December 31, 2019 (see \cite{eckert1990property}). The samples used for housing market estimation are not necessarily random draws from the population of dwellings, but are all the properties recently sold in the city of Boulder. \cite{robinson1979housing} states that \qq{recent sales are not necessarily a random draw from the total housing stock. If the purpose is to index the market of available units, this may not be of great concern, but if the purpose is to index the total stock, we must concern ourselves with possible selection bias}. Several studies have tested the existence of such biases. So far most studies have found the magnitude of the bias to be modest (see e.g., \cite{gatzlaff1998sample}).\\

We collected real estate datasets from different resources consisting of Multiple Listing Service databases\footnote{\url{https://realtyna.com/blog/list-mls-us}}, Public School Ratings\footnote{\url{https://www.greatschools.org}}, Colorado Crime Rates and Statistics Information\footnote{\url{https://www.neighborhoodscout.com/co/crime}}, CrimeReports\footnote{\url{https://www.cityprotect.com}}, WalkScore\footnote{\url{https://www.walkscore.com}}, and US Census Bureau\footnote{\url{https://data.census.gov/cedsci}}. Before proceeding with the estimation of our models we merged all data sets obtained from these websites. We screened our collected data set. Among these transactions 4 houses were in rough shape and needed to be rebuilt. Consequently, we deleted those 4 observations. All houses except one; which has lots of luxury furniture, have been sold unfurnished. So, we dropped the only furnished property. 30 properties were reported as horse properties. We eliminated those associated records. We checked for duplicate records. We found 4 duplicate transactions and eliminated those. In a number of observations we faced missing data points for some characteristics. For example, the number of bedrooms, bathroom, parking, $\text{Lot Area}$, HOA fees, $\text{Solar Power}$ and Pool, Bath tub, Sauna, or Jacuzzi were missing in some records. We updated some of these missing data points by visiting different websites and looking up for the correct information. Finally, a few observations left with missing data points for the continuous variable \text{Lot Area} and the dummy variables $\text{Solar Power}$ and Pool, Bath tub, Sauna, or Jacuzzi. To deal with missing data points we could eliminate from the model any attributes that have missing values, or alternatively we could drop all those properties with missing attributes. However, deleting observations due to incomplete list of attributes can cause sample size reduction and sample selection bias (see e.g., \cite{hill2013hedonic}). Instead, to avoid losing the samples from analysis we imputed missing values in the \qq{lot-size} variable with its mean. Similarly, we imputed missing values for the dummy variables $\text{Solar Power}$ and Pool, Bath tub, Sauna, or Jacuzzi with their mode. Another problem we encountered is the presence of outliers. The outliers were identified by calculating the $95\%$ percentile. To mitigate the influence of outliers on the analysis, we applied one-sided $95\%$ or two-sided $90\%$ Winsorization. That means outliers were set to the value of the $95\%$ percentile. The data cleaning process left us with a final sample of $1018$ observations. Information about the descriptive statistics of the variables in this data set can be found in Table \ref{Table1} and \ref{Table2}. In addition to the mean, standard deviation, minimum and maximum, information about the relative standard deviation (the coefficient of variation ($CV=\frac{\sigma}{\mu}$), which represents the extent of variability in relation to the average of the variable has been reported in Table \ref{Table1}. \\

\begin{table}[h!]
\caption{Descriptive Statistics for Numerical Variables.}\label{Table1}
\centering
\begin{adjustbox}{max width=\textwidth}
\begin{tabular}{@{}l l l l l l@{}}\hline\hline\toprule\\
\multirow{2}[3]{*}{Variables} & \multicolumn{5}{c}{Citywide Level}  \\\\
\cmidrule(lr){2-6}\\

&Mean & St. Dev.&Min &Max & Coefficient of Variation (CV) \\ [0.5ex] % 
\hline
\hline
\\
$\text{House Price}$ ($\$$)& $896,332$ & 679,195.8 & 112,897 & 7,200,000 &$76\%$\\

% $\text{Ln} (\text{House Price})$& 13.49 & 0.57 & 12.18 & 14.57&$4\%$\\

% $\text{Ln} (\text{Lot Area})$ (SqFt) & 6.21 & 4.37 & 0.00 & 14.27 & 70\%\\

% Ln(Living Area) (SqFt)& 7.54 & 0.61 & 6.03 & 9.25  &8\%\\

$\text{Lot Area}$ (SqFt)& 18,367 & 84,909.71 & 0 & 1,577,744   &462.29\%\\

$\text{Living Area}$ (SqFt)& 2,264 & 1,398.79 & 416 & 10,354   &61.78\%\\

Age (year) & 43.12 & 21.46 & 1 & 98&50\%\\

%Number of Bedrooms  & 3.16 & 1.17 & 0 & 7& Numerical \\

Number of Full Bathrooms  & 1.55 & 0.76 & 0 & 3& 49\% \\

Number of Half Bathrooms  & 0.41 & 0.51 & 0 & 2& 124\%\\ 

Number of $\frac{3}{4}$ Bathrooms  & 0.64 & 0.69 & 0 & 2& 108\%\\

Parking & 1.68 & 0.71 & 0 & 3& 42\%  \\ 

HOA Fees (annually) ($\$$)  & 1,693.32 & 2,033.16 & 0 & 7,113& 120\%\\ 

Drive to CBD (minutes)  & 11.42 & 6.91 & 1 & 26& 61\% \\

Walk to E.School (minutes) & 21.37 & 17.31 & 2 & 68& 81\% \\ 

Walk to M.school (minutes) & 33.21 & 27.72 & 2 & 96  & 83\%\\ 

Walk to H.school (minutes) & 46.94 & 32.92 & 4 & 122 & 70\%\\ 

Married ($\%$) & 42.97 & 16.87 & 9.90 & 70.30& 40\%\\

Median Household Inc. ($\$$)  & 61,137.44 & 20,891.56 & 19,985 & 96,406&  34\%\\

Neighborhood's Population & 43,641.85 & 46,872.42 & 888 & 99,081 & 107\%\\\\
\hline

Sample size  & 1018   \\ 
 [1ex]

\hline \hline
\end{tabular} %You can adjust how far below the table the text should appear
\end{adjustbox}
\label{table:nonlin}
\caption*{\textbf{Note}: Homeowner's association (HOA). Central business district (CBD).~~~~~~~~}
\end{table}

\begin{table}[!hp]
\caption{Descriptive Statistics for Categorical Variables.}\label{Table2}
\centering
\begin{adjustbox}{max width=\textwidth}
\begin{tabular}{@{}l l l l  l@{}}\hline\hline \toprule\\
\multirow{2}[3]{*}{Variables} & \multicolumn{4}{c}{Citywide Level}  \\\\
\cmidrule(lr){2-5}\\

& Levels& Description & Frequency & Percent \\ [0.5ex] % inserts table %heading
\hline
\hline\\
& 0 &No& 658 &64.64 \\[-1ex]  

\raisebox{1ex}{Pool, Bath tub, Sauna, or Jacuzzi} &  1&Yes&360&35.36 \\[1.5ex]

& 0&No & 724 &71.12\\[-1ex]  

\raisebox{1ex}{Solar Power} &  1&Yes &294&28.88\\[1.5ex]
 
& 1&A & 334 &32.81  \\[-1ex]  

\raisebox{1ex}{Nearest E.School Rank} & 2  &B
&548 &53.83  \\

& 3 &C &136 &13.36\\[1.5ex]
 
& 1 &A& 125&12.28 \\[-1ex]  

\raisebox{1ex}{Nearest M.School Rank } & 2  &B
&633 &62.18 \\

& 3& C &260 &25.54 \\[1.5ex]
 
\raisebox{1ex}{Nearest H.School Rank} & 1 &A
&1018 &100 \\[1.5ex]

& 1&Central & 230&22.59 \\[.01ex] 

& 2&North 
&238 &23.38  \\[-.05ex] 
& 3& South &143 &14.05 \\[-1ex] 
\raisebox{1ex}{Region} & 4& East&200 &19.65 \\

& 5& Gunbarrel&125 &12.28 \\

& 6& Rural&82 &8.06 \\[1.5ex]

& 0 &0 bedroom & 3&0.29\\[.01ex] 

& 1 &1 bedroom & 72&7.07\\[-.05ex] 

& 2&2 bedrooms &253 &24.85 \\ [-1ex] 

\raisebox{1ex}{Number of Bedrooms} 
& 3& 3 bedrooms&285 &28.00 \\

& 4& 4 bedrooms&249 &24.46 \\

& 5& 5 bedrooms&127 &12.48 \\

& 6& 6 bedrooms&27 &2.65 \\

& 7& 7 bedrooms&2 &0.20 \\[2.4ex]

& 1 &Condominum & 324&31.83 \\[-1ex]  

\raisebox{1ex}{Property Type} & 2&Town - Home 
&90 &8.84  \\

& 3& Single Family &604 &59.33 \\[1.5ex]

& 1 &Highest crime rate&82&8.06\\[-.75ex]  

\raisebox{1ex}{Neighborhood's Crime Level} & 2  &Middle crime rate&505&49.61\\

& 3& Lowest crime rate & 431 &42.34\\[1.75ex]
 
 \hline
Sample size&-&1018&100&-\\

\hline \hline
\end{tabular} %[2.3] %You can adjust how far below the table the text should appear
\end{adjustbox}
\label{table:nonlin}
\end{table}

\begin{figure}[t]
 \makebox[\textwidth][c]{\includegraphics[width=19cm,height=0.52\textheight]{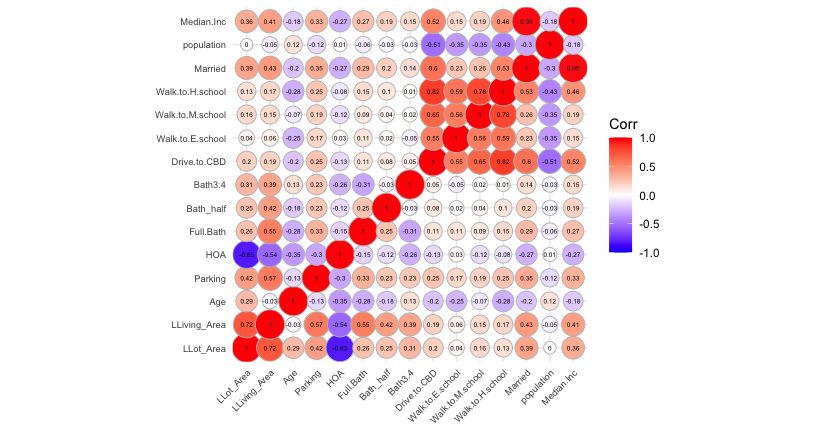}}
\caption{Correlation Plot.}\label{Figure5}
\end{figure}

Before performing our models we should test for multicollinearity problem. Figure \ref{Figure5} summarizes the Pearson correlation coefficients. Positive correlations are shown in red color and negative correlations in dark blue color. The color intensity and the size of the circle are proportional to the correlation coefficients. We drop variables which generate pairwise correlation coefficients greater (smaller) than the common thresholds $0.8$ (-$0.8$). We also calculate the generalized variance inflation factors (GVIF). The accepted GVIF cutoff is $\sqrt{VIF}=\sqrt{5}=2.24$.\\  

We standardize the numerical variables in our data set. In order to standardize the variables we first compute the average value and the standard deviation of each variable in our data set. Next, we subtract the mean from each variable and divide it by its standard deviation. As a result, the mean value of each numerical variable is approximately zero, and the standard deviation is almost one. Standardizing the data usually speeds up learning and leads to more efficient, faster convergence, and better predictions.\\

%--It would be problematic to feed into a neural network values that all take wildly different ranges. The network might be able to automatically adapt to such heterogeneous data, but it would definitely make learning more difficult. A widespread best practice to deal with such data is to do feature­wise normalization: for each feature in the input data (a column in the input data matrix), you subtract the mean of the feature and divide by the standard deviation, so that the feature is centered around 0 and has a unit standard deviation. This is easily done in R using the scale() function.

The ANN and KNN algorithms would accept only numeric data and cannot directly work with categorical variables. Consequently, we apply the one-hot encoding method to convert the categorical variables into numbers. The one-hot encoding for a variable with $n$ categories creates $n$ new binary variables, one for each possible category and assigns values of one and zero.\\ 

Usually, we are interested in how well the deep learning and machine learning algorithms perform on data that it has not seen before, since this determines how well it will work when deployed in the real world. We therefore evaluate these models' performance using a test data which is separate from the data used for training the deep learning and machine learning models.\\

To evaluate the predictive accuracy and reliability of our models, we split our sample into training, validation, and test data sets. In this study, the training data set is a sample of $80\%$ random observations from the $1018$ total transactions. We train the model over the training data set. These transactions are used to estimate the parameters of the models. The remaining $20\%$ random samples are kept out from the training data set and are evenly divided into validation and test data sets. Validation data set is used to tune the hyperparameters and the test data set is used to assess prediction precision of the models.\\

\begin{figure}[h!]
\centering
\hspace*{.00001\textwidth} 
\includegraphics[width=0.95\textwidth]{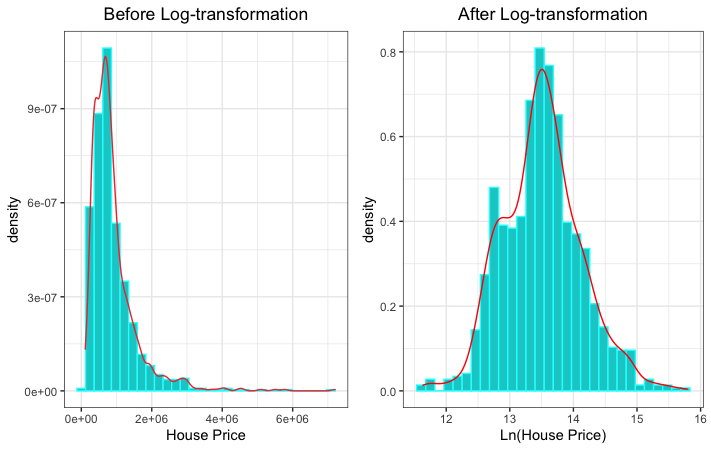}
\caption{Distribution of House Prices before and after Applying the Log-transformation.}\label{Figure6}
\end{figure}

To process the data, we are doing feature engineering. We apply the log-transformation on the house prices. The log-transformation of the dependent variable (house prices) often mitigates the problem of heteroskedasticity associated with the use of the highly skewed sales price variable. Figure \ref{Figure6} plots the distribution of house prices before and after applying the log-transformation. In addition, we are adding categorical variables in our models to control for location and property differences.

\section{Results}

In this study, all of the models were implemented using the open-source R statistical computing environment with Keras, randomForest, and Caret packages (see e.g., \cite{rcolorbrewer2018package}). We run these models separately to develop housing price prediction models. To evaluate the performance of these models we computed the performance metrics, namely root mean squared error (RMSE), mean absolute percentage error (MAPE), and ${\text R^2}$.\\

We first make use of linear regression model to estimate the hedonic price equation over the training data set, at the citywide market level. The housing market in the aggregate market level contains high variation in the houses attributes. We both graphically and statistically detected heteroskedasticity. In the presence of heteroskedasticity OLS estimators are unbiased, while the standard error and the statistics we used to test hypotheses based on ordinary least square estimation method will be inefficient. Following \cite{fletcher2000heteroscedasticity} we use White’s standard errors test to correct the issue of incorrect standard errors and test statistics. Moreover, to control for property type and location differences we are adding categorical variables into the hedonic pricing model. The robust standard errors and heteroskedasticity-consistent estimates are reported in Table \ref{Table3}.\\ 

\begin{table}[!h]
\caption{Hedonic Model Estimates - White Correction.\\}\label{Table3}
\centering %used for centering table
\ra{1.3}
\begin{adjustbox}{max width=\textwidth}
\begin{tabular}{@{}lllllrr@{}}\toprule
\hline
\multirow{2}[3]{*}{Independent Variables} & \multicolumn{3}{c}{Citywide Level} \\ \cmidrule{2-4} 
& Coeff. & White’s Std. Error& White’s T-Statistics\\ \midrule
\hline\\
(Intercept) & $1.21$ & $0.25$ & $4.93^{***}$ \\ 

Ln(Lot Area) & $0.81$ & $0.15$ & $5.38^{***}$ \\

%$(\text{Ln} (\text{Lot Area}))^2$ & -$0.004$ & $0.005$ & -$0.85$ \\ 

Ln(Living Area) & $0.56$ & $0.04$ & $13.89^{***}$  \\

%$(\text{Ln} (\text{Living Area}))^2$ & $0.07$ & $0.03$ & $2.34^{**}$ \\

Age & -$0.05$ & $0.02$ & -$2.37^{**}$ \\

%$\text{Age}^2$ & $0.0001$ & $0.0000$ & $5.73^{***}$ \\

\textbf{Number of Bedrooms}\\

(Bedrooms)2 & $0.18$ & $0.07$ & $2.61^{***}$ \\

(Bedrooms)3 & $0.13$ & $0.08$ & $1.55$ \\ 

(Bedrooms)4 & $0.04$ & $0.09$ & $0.42$\\ 

(Bedrooms)5+ & -$0.04$ & $0.10$ & -$0.34$\\ 

Number of Full Bathrooms & $0.11$ & $0.02$ & $4.48^{***}$ \\

Number of Half Bathrooms & -$0.002$ & $0.02$ & -$0.13$ \\ 

Number of $\frac{3}{4}$ Bathrooms & $0.07$ & $0.02$ & $3.42^{***}$\\

Parking & $0.04$ & $0.02$ & $2.17^{**}$ \\ 

HOA Fees & $0.04$ & $0.04$ & $0.94$\\

\textbf{School Ranking}\\

(Nearest E.School Rank)2 & -$0.06$ & $0.04$ & -$1.33$ \\

(Nearest E.School Rank)3 & -$0.36$ & $0.06$ & -$5.64^{***}$ \\ 

(Nearest M.School Rank)2 & -$0.32$ & $0.09$ & -$3.64^{***}$\\ 

(Nearest M.School Rank)3 & -$0.21$ & $0.11$ & -$1.98^{**}$ \\ 

(Pool, Bath tub, Sauna, or Jacuzzi)1 & $0.004$ & $0.04$ & $0.11$ \\

(Solar Power)1 & $0.18$ & $0.07$ & $2.60^{***}$ \\

Drive to CBD & -$0.26$ & $0.04$ & -$6.06^{***}$ \\

Walk to E.School & -$0.0001$ & $0.02$ & -$0.01$ \\ 

Walk to M.school & -$0.01$ & $0.04$ & -$0.30$\\ 

Median Household Inc. & -$0.01$ & $0.03$ & -$0.22$\\

Neighborhood's Population& -$0.02$ & $0.02$ & -$0.97$\\

\hline \hline
\end{tabular}
\end{adjustbox}
\end{table}

\begin{table*}[t]
\ContinuedFloat
\caption{(Continued.)}
 %used for centering table
\ra{1.3}
\begin{adjustbox}{max width=\textwidth}
\begin{tabular}{@{}llllcrrr@{}}\toprule
\hline
\multirow{2}[3]{*}{Independent Variables} & \multicolumn{3}{c}{Citywide Level} \\ \cmidrule{2-4} 
& Coeff. & White’s Std. Error& White’s T-Statistics\\ \midrule
\hline

\textbf{Crime Levels} (Base: Level 3) \thickspace  \thickspace\\

Neighborhood's Crime Level: 1 & -$0.20$ & $0.07$ & -$2.88^{***}$\\ 

Neighborhood's Crime Level: 2 & -$0.06$ & $0.04$ & -$1.55$ \\

\textbf{Property Type}\\

Single Family Houses & -$0.97$ & $0.31$ & -$3.14^{***}$ \\

Town-Home Houses & -$1.04$ & $0.27$ & -$3.80^{***}$\\

\textbf{Regions} (Base: Central)\\

(North)2 & -$0.30$ & $0.07$ & -$4.11^{***}$  \\ 

(South)3 & -$0.30$ & $0.10$ & -$3.11^{***}$\\ 

(East)4 & -$0.44$ & $0.08$ & -$5.53^{***}$\\ 

(Gunbarrel)5 & -$0.43$ & $0.14$ & -$3.16^{***}$ \\ 

(Rural)6 &  -$0.64$ & $0.14$ & -$4.55^{***}$\\ \\

\hline

Chi-squared ($p-$value)& $105$&($0.0009$)\\
Dependent variable: Ln(Price) \\\bottomrule
\hline \hline
\end{tabular}
\end{adjustbox}
\caption*{\textbf{Note}: $***,~ **$, and $*$ ~indicate $p < 0.01$, $p < 0.05$, and $p < 0.1$ respectively.\\ Base group in the citywide housing market: condominium houses, with one bedroom or less, no pool, bath tub, sauna, or jacuzzi, no solar power, in central Boulder neighborhood with crime level 3. }
\end{table*}

Next, we perform the deep learning and machine learning algorithms, ANN, RF, and kNN, to predict the house prices in the city of Boulder, Colorado. The structure of deep learning and machine learning algorithms, will substantially affect the performance of the models' predictions. The optimal structure is the one that corresponds to the lowest test error rate. However, using the test data set for hyperparameter tuning can lead to overfitting. As such, to evaluate our model performance we use validation data set to estimate the test error. In order to find an accurate and efficient estimation for the target variable we find the values for the hyperparameters by minimizing the mean squared error (MSE) for our validation data sets. This allows us to obtain a high percentage of variance explained and avoid model’s over-fitting. We determine the best structure of our models using the validation data by a random (grid) search method. Following a random search method we test different model structures on the validation data set in order to select the best performing model.\\

For instance in the ANN model, depth and width; the number of layers (depth) and the number of nodes (width) that are included in each layer will substantially affect the performance of learning and prediction of the network. In RF, different values of the hyperparameters namely, number of trees, number of input variables drawn randomly for each node (mtry), the splitting rule, and the maximum depth provide different outcomes. In the kNN approach, the predicted value for the target variable depends on the value of the hyperparameter $k$; the number of neighbors. The optimal hyperparameters are those that result in the lowest test error rate. By holding out the validation data set we are estimating the test error rate. As such, we tune the hyperparameters in these models. \\

To find the optimal values of the hyperparameters we randomly selected training, validation, and test data sets for $20$ times. We ran ANN, RF, and kNN models over all these samples and found the optimal hyperparameters over all these 20 samples separately. Next, we averaged all these optimal hyperparameters. After tuning the hyperparameters we fixed our model structures by the optimal values of the hyperparameters. More details on the optimal values of these hyperparameters over our data set are shown in the Appendix A.\\

To appraise the predictive performance of all these models in predicting housing prices on the unseen (test) data set, given the optimal values of the hyperparameters over these $20$ samples we used some performance metrics such as root mean squared error (RMSE) that capture the distance between predicted and observed transaction price, mean absolute error (MAE), mean absolute percentage error (MAPE), that captures the average percentage deviation of predictions from the actual transaction prices, and ${\text R^2}$ that measures the explanatory power of these models. These performance metrics are defined as:\\

$RMSE= \sqrt {{\frac {\sum _{i=1}^n{(y_i -\hat{y_i})}^2}{n}}}$\\

$MAPE= \frac{100}{n} \times \sum _{i=1}^n|{\frac{y_i -\hat{y_i}}{{y_i}}}|$\\

$R^2=1-\frac{\sum _{i=1}^n{(y_i -\hat{y_i})}^2}{\sum _{i=1}^n{(y_i -\Bar{Y})}^2} $\\

\noindent where ${y_i}$ is the actual price value, $\hat{y_i}$ is the predicted value, n is the number of samples, and $\Bar{Y}$ is the average of the target variables.\\

We calculated all these performance metrics (RMSE, MAE, and ${\text R^2}$) over all these 20 samples and averaged these values. The results are reported in \ref{Table4}. Among HP, ANN, RF, and KNN models, we observe that RMSE and MAPE are the smallest in the RF model. The performance measures RMSE and MAPE in the ANN model are larger than in RF but smaller than in kNN and HP models. RMSE and MAPE are lower in HP model than in the kNN model. To examine whether or not these differences are statistically significant we performed a paired T-test over these 20 samples. We observed that these differences are statistically significant at $1\%$ significance level. See Table \ref{Table4} for the details.\\

\begin{sidewaystable} 
\setlength\tabcolsep{0pt}
\caption{Comparison of the performance of HP model with ANN, RF, and kNN.}\label{Table4}
\centering %used for centering table
\ra{1.3}

\begin{adjustbox}{max width=\textwidth}
\begin{tabular}{@{}lllllllllllllllllll@{}}\toprule
\hline
\multirow{2}[5]{*}{Measurement} \thickspace \thickspace \thickspace \thickspace& \multicolumn{3}{c}{HP} & \phantom{abc}& \multicolumn{3}{c}{ANN} &
\phantom{abc} & \multicolumn{3}{c}{RF}&\phantom{abc} & \multicolumn{3}{c}{kNN}\\ \cmidrule{2-4} \cmidrule{6-8} \cmidrule{10-12} \cmidrule{14-16}
& Training \thickspace \thickspace & Validation\thickspace \thickspace& Test&& Training \thickspace \thickspace& Validation \thickspace \thickspace& Test && Training \thickspace \thickspace & Validation \thickspace \thickspace& Test&& Training \thickspace \thickspace& Validation \thickspace \thickspace& Test\\ \midrule
\hline\\

RMSE & $0.22$ $(0.003)$ \thickspace \thickspace & $0.23$ $(0.02)$\thickspace \thickspace & $0.24$ $(0.01)$ \thickspace \thickspace&& $0.04$  $(0.02)$\thickspace \thickspace& $0.21$ $(0.02)$ \thickspace \thickspace& $0.22$  $(0.02)$ \thickspace \thickspace&& $0.09$ $(0.002)$ \thickspace \thickspace & $0.21$ $(0.02)$ \thickspace \thickspace& $0.21$ $(0.03)$ \thickspace \thickspace&& $0.22$  $(0.03)$ \thickspace& $0.27$ $(0.03)$ \thickspace \thickspace& $0.28$ $(0.03)$  \\ 

MAEP & $1.21\%$ & $1.26\%$ & $1.36\%$&&$0.2\%$ &$1.12\%$ & $1.21\%$&& $0.48\%$&$1.08\%$&$1.14\%$ &&$1.2\%$&$1.44\%$&$1.58\%$ \\

$\text{R}^2$  && &$0.82$ \thickspace \thickspace&  &&& $0.85$&&&& $0.86$&&&& $0.75$ &  &   \\ \\\bottomrule
\hline \hline
\end{tabular}
\end{adjustbox}
\caption*{ \textbf{Note}: The standard deviations of these performance metrics are represented in the parenthesis.~~~~~~~~~~~~ ~~~~~~~~~~\thickspace \thickspace \thickspace \thickspace \thickspace \thickspace\thickspace \thickspace \thickspace \thickspace \thickspace \thickspace \thickspace \thickspace \thickspace \thickspace\\ }
\end{sidewaystable} 

To compare the explanatory power of these models we also compare the ${\text R^2}$ values of these models. Among these models the ${\text R^2}$ values ranged between $(0.75, 0.86)$. The ${\text R^2}$ of the RF model is $0.86$, which implies that $86\%$ of the variation of the dependent variable has been explained by the available features in the model. This value is the highest amongst these models.\\

\begin{figure}
 \centering
 \makebox[\textwidth][c]{\includegraphics[width=11cm]{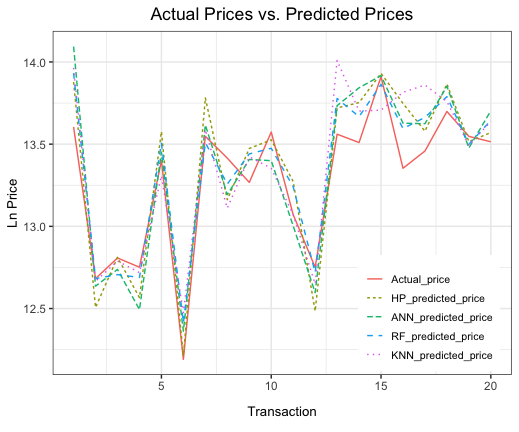}}
 \caption{Comparison of the Performance of the HP, ANN, RF, and kNN Models.}\label{Figure7}
 \end{figure}

Additionally, for the purpose of comparison between different models performance the predicted prices of the first 20 properties obtained by these models were compared with the actual prices. Figure \ref{Figure7} depicts the estimated house prices obtained by these models versus the actual values. This graph represents the RF price predictions are closer to the purchase prices than other models' predictions. \\

Given the value of RMSE, MAPE, and ${\text R^2}$ and visualizing the actual and predicted prices through differen models we realize that RF model outperformed all other models in this housing market. ANN is the second best performing method. HP model performed better than the kNN model.\\

Now we aim to learn what features have the biggest impact on house prices in the city of Boulder, Colorado. We investigate the $22$ variables in Table \ref{Table1} and \ref{Table2} to determine which of them have a significant or dominant impact on the house prices. We rely on RF model which is a popular method for feature ranking. We measure the importance of a feature by calculating the increase in the model's prediction error after permuting the feature (see e.g., \cite{breiman2001random} and \cite{fisher2019all}).\\

\begin{figure}[h!]
\centering
\hspace*{.0001\textwidth} 
\includegraphics[width=0.95\textwidth]{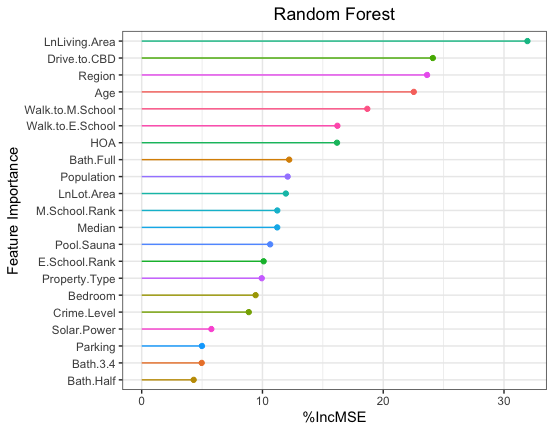}
\caption{Feature Importance - Random Forest Model. }\label{Figure8}
\end{figure}

Figure \ref{Figure8} shows the importance of the variables in the trained RF model. The increase in the values of the MSE, are depicted in this figure. The RF model selects the \qq{$\text{Ln} (\text{Living Area})$} as the most important predictive variable for the property prices in the city of Boulder. The \qq{Drive to CBD} is the second most important variable followed by \qq{Region} and \qq{Age}. The variable \qq{Bath.Half} (the half of bathroom) is the least important feature.\\

\begin{figure}[h!]
\centering
\includegraphics[width=0.9\textwidth]{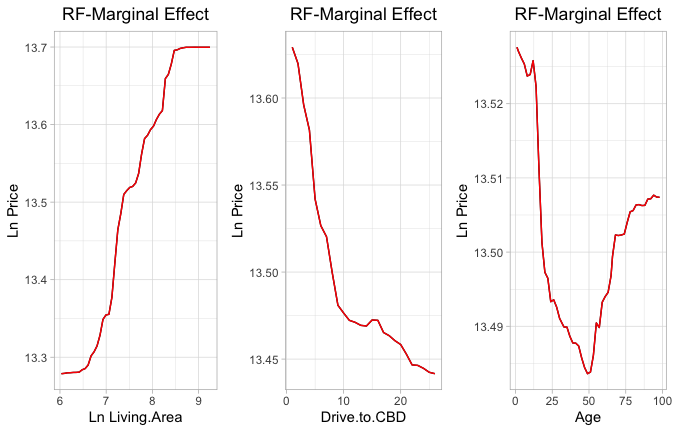}
\caption{Marginal Effect of the Three Most Important Numerical Features on the House Value from the Random Forest Model. }\label{Figure9}
\end{figure}

Figure \ref{Figure9} shows the marginal effect of the three most important features on the property values in the city of Boulder, Colorado. All these variables clearly show non-linear association with the target variable.  We observe that the feature $ (\text{Living Area})$ has positive marginal effect on the property values. As $ (\text{Living Area})$ increase the house price increases at a different rate. As the residential property distance to CBD increases the house value decreases at a non-constant rate. Dwellings built in during the last $50$ years have negative marginal effect on house prices. However, houses with the age above $50$ years have positive marginal effect on property values.\\

Since the marginal effects in Figure \ref{Figure8} depicted non-linear patterns, the machine learning and deep learning techniques such as RF and ANN can be considered as more accurate and efficient prediction methods for the real estate market prices in the city of Boulder instead of hedonic price regression model.\\

\section{Conclusion}

% The objective of this study was to develop property appraisal models in the city of Boulder, Colorado by implementing hedonic price regression model, deep learning and machine learning techniques.

% In this study we compared the predictive power of the HP model with ANN, RF, and kNN models for the house values. We observed non-linear association between the house features and house prices. As such, the machine learning and deep learning techniques can be considered as more accurate and efficient prediction methods for the real estate market prices in the city of Boulder that hedonic price regression models.\\

In this paper we implemented hedonic price regression model, deep learning and machine learning techniques to develop property appraisal models in the city of Boulder, Colorado. In this study we compared the predictive power of the HP model with ANN, RF, and kNN models for the house values. We observed non-linear association between the house features and house prices. As such, the machine learning and deep learning techniques can be considered as more accurate and efficient prediction methods for the real estate market prices in the city of Boulder than hedonic price regression models.\\

We compared several performance measurements for the predictions of the house prices in the unseen (test) data set. Considering the ${R^2}$, RMSE and MAPE values for these approaches, it can be concluded that the RF outperforms the other methods in this housing market. It seems the RF method captures the complexity or non-linearity of the housing markets in the city of Boulder more than other models.\\

In general, when there is small data set random forest is a good algorithm, since it is a bagging of decision trees with bootstrap. Each decision tree is fed with a sample of data with replacement, in this way even if the data set is small there are bigger chances of making a good model. Moreover, RF can handle categorical variables with many levels. In the case of ANN method, for categorical variables with many levels the one-hot encoding creates a large number of binary variables which lead to a larger number of estimated parameters and this usually results in overfitting.\\

The most important aspect of the deep learning and machine learning approaches are that, they learn directly from observed data and there is no need to assume explicit functions between the inputs and output of the studies. They have the ability to rearrange their structure by adjusting the values of parameters and hyperparameters. \\

As dwelling price prediction involves many non-linear variables, in order to uncover non-linear relationships between the property attributes and predict the value of a house the deep learning and machine learning algorithms should be considered. Deep learning and machine learning algorithms such as ANN and RF are accurate and efficient prediction methods. They have however, a serious limitation and that is they are not interpretable. In contrast, the coefficients in the hedonic models are easily interpretable.\\

There are however, some methods such as Lime, Shap and Eli5 that make the deep learning and machine learning algorithms interpretable. In future studies it would be useful to use those methods. It would be also interesting to expand our study and in addition to the micro factors such as neighborhood, location, and structure characteristics consider macroeconomic variables such as unemployment rate. A more accurate forecast of the real estate market prices must exploit not only variables relating to characteristics of the market, but also combine them with different information sources such as macroeconomic factors.  \\

\section*{Acknowledgements}
Thanks goes to Dr. Nicholas E. Flores\footnote{Department Chair of Economics, Colorado University at Boulder} for his comments.

\newpage
\renewcommand\bibname{References}

%\section*{Bibliography}
\bibliographystyle{apalike}
%\bibliography{referencesfile}

%\bibliographystyle{apalike}
\bibliography{mybib.bib}
\newpage

\clearpage
\pagenumbering{arabic}% resets `page` counter to 1
\renewcommand*{\thepage}{A\arabic{page}}
\renewcommand\appendixtocname{Appendices}
\clearpage
\appendix
\renewcommand\thefigure{\thesection.\arabic{figure}} 
\renewcommand\thetable{\thesection.\arabic{table}}

\label{app:Online}
%\section{appfoo}

\begin{appendix}
\section{Appendix}

\begin{table}[ht]
\caption{Description of Models’ Parameter Tuning.}\label{Table3}
\captionsetup{singlelinecheck = false}
\centering
\begin{adjustbox}{max width=\textwidth}
\begin{tabular}{llllllllllllll}
\hline\hline\\
Models && Parameter Tuning on average & Reference \\ [0.5ex] % inserts table %heading
\hline \hline
\\

\raisebox{.05ex}{ANN} 
&& Units: 192& Grid search \\

&& Activation: RELU & Grid search  \\

&& Layer Dense: 2& Grid search \\

&& Kernel Regularizer: regularizer\_l2 (0.001)& Grid search \\
&&Epochs: 100000 & Grid search\\
&&Batch Size: & Grid search\\
%– Batch Size = 2048;
&&Learning Rate: 0.001 & Grid search\\
&&Optimizer: Adam optimizer &\\
&&Loss Function: MSE &\\[2.4ex]

%– Dropout Rate = 0.1.

\raisebox{0.01ex}{RF} 
 &&Ntree:250& Grid search \\
 &&Mtry:7& Grid search\\[0.4ex]
\\
kNN &&K:7&Grid search\\\\[1ex]

\hline \hline
\end{tabular}
\end{adjustbox}
\label{table:nonlin}

\end{table}

\end{appendix}

\end{document}